\documentclass[9pt,twocolumn,twoside]{pnas-new}
\setboolean{displaywatermark}{false}

\templatetype{pnasresearcharticle} 

\title{\textbf{Coherent detection of hidden spin-lattice coupling in a van der Waals antiferromagnet}}

\author[1,*]{Emre Ergeçen}
\author[1,*]{Batyr Ilyas}
\author[2,3]{Junghyun Kim}
\author[2,3]{Jaena Park}
\author[1]{Mehmet Burak Yilmaz}
\author[1]{Tianchuang Luo}
\author[4,5]{Di Xiao}
\author[6]{Satoshi Okamoto}
\author[2,3]{Je-Geun Park}
\author[1,$\dag$]{Nuh Gedik}

\affil[1]{Department of Physics, Massachusetts Institute of Technology, Cambridge, 02139, Massachusetts, USA.}
\affil[2]{Center for Quantum Materials, Seoul National University, Seoul 08826, Republic of Korea.}
\affil[3]{Department of Physics and Astronomy and Institute of Applied Physics, Seoul National
University, Seoul 08826, Republic of Korea.}
\affil[4]{Department of Materials Science and Engineering, University of Washington, Seattle, Washington, USA}
\affil[5]{Department of Physics, University of Washington, Seattle, Washington, USA}
\affil[6]{Materials Science and Technology Division, Oak Ridge National Laboratory, Oak Ridge, Tennessee 37831 USA}

\leadauthor{Ergeçen} 

\significancestatement{With the advent of van der Waals (vdW) materials, emergent phases of matter started to themselves themselves in two-dimensional devices. In particular, vdW magnets show strong interactions between lattice and spins, leading to diverse magnetic ground states. Our work provides a definitive mechanism to spin-lattice coupling in vdW magnets, FePS$_3$ and NiPS$_3$, by employing phase resolved coherent phonon spectroscopy. We observe sudden change in the phase of a specific phonon, undetectable by phase-insensitive conventional techniques, such as Raman and X-ray scattering. Our work directly establishes the role of trigonal distortions and d-orbital occupation in spin-lattice coupling in vdW magnets.}

\authorcontributions{E.E., B.I. and M.B.Y. built the optical spectroscopy setups. E.E. and B.I. performed the experiments. E.E., B.I. and N.G. analysed and interpret the data with crucial inputs from T.L. E.E. developed the theoretical model under the supervision of D.X. and S.O. J.K. and J.P. synthesized and characterized single crystals of FePS$_3$, supervised by  J.-G.P. E.E., B.I. and N.G. wrote the manuscript with crucial inputs from all authors. This project was supervised by N.G.}
\authordeclaration{The authors declare no competing interests.}
\equalauthors{\textsuperscript{*}E.E.(Author One) contributed equally to this work with B.I. (Author Two).}
\correspondingauthor{\textsuperscript{$\dag$}To whom correspondence should be addressed. E-mail: gedik@mit.edu}


\begin{abstract}
Strong interactions between different degrees of freedom lead to exotic phases of matter with complex order parameters and emergent collective excitations. Conventional techniques, such as scattering and transport, probe the amplitudes of these excitations, but they are typically insensitive to phase. Therefore, novel methods with phase sensitivity are required to understand ground states with phase modulations and interactions that couple to the phase of collective modes. Here, by performing phase-resolved coherent phonon spectroscopy (CPS), we reveal a hidden spin-lattice coupling in a vdW antiferromagnet FePS$_{3}$ that eluded other phase-insensitive conventional probes, such as Raman and X-ray scattering. With comparative analysis and analytical calculations, we directly show that the magnetic order in FePS$_{3}$ selectively couples to the trigonal distortions through partially filled t$_{2g}$ orbitals. This magnetoelastic coupling is linear in magnetic order and lattice parameters, rendering these distortions inaccessible to inelastic scattering techniques. Our results not only capture the elusive spin-lattice coupling in FePS$_3$, but also establish phase-resolved CPS as a tool to investigate hidden interactions.
\end{abstract}

\begin{document}

\maketitle
\thispagestyle{firststyle}
\ifthenelse{\boolean{shortarticle}}{\ifthenelse{\boolean{singlecolumn}}{\abscontentformatted}{\abscontent}}{}

\dropcap{V}an der Waals (vdW) materials and their heterostructures host tunable correlated phenomena, which arise from the interplay between lattice, spins and orbitals, ranging from unconventional superconductivity \cite{Cao2018} to charge-ordered states \cite{Xu2020}. In particular, vdW magnets have recently emerged as platforms to study strong interactions between lattice and spins \cite{Gibertini2019,Huang2020,Burch2018}. For example, the stacking configuration of two adjacent layers can tune the interlayer exchange interaction \cite{sivadas2018stacking}, leading to diverse magnetic ground states \cite{Huang2017}, such as the coexistence of two distinct magnetic states with a mesoscopic periodicity defined by the twist angle \cite{Song2021, Xu2021, Xie2021}. 

Despite a plethora of phases that emerge from strong spin-lattice coupling in vdW magnets, their microscopic origins are not well understood and demand new techniques with magnetic and structural sensitivity. This is especially true for FePS$_{3}$ (Figure 1a), a prototypical vdW antiferromagnet, which exhibits easy-axis zigzag antiferromagnetic order below its N\'eel temperature (T$_\textrm{N}$=118 K) \cite{LeFlem1982}. FePS$_{3}$ displays an unusual set of magnetically active low energy collective modes \cite{Lee2016, Lancon2016, Scagliotti1987, Ghosh2021}, interpreted as zone folded phonon modes, indicating strong spin-lattice coupling. However, the structural phonon modes that exist above the N\'eel temperature do not experience any frequency renormalization, a hallmark of strong spin-phonon coupling \cite{Hong2012, Vermette2008}, despite the emergence of new magnetically enabled phonons at the onset of magnetic order. The microscopic mechanisms leading to the dichotomy between new collective modes and structural phonon modes, and the nature of spin-lattice coupling in FePS$_{3}$ still remain an open question.

In this work, we employ phase-resolved coherent phonon spectroscopy (CPS) to reveal a mode selective spin-lattice coupling in FePS$_{3}$ that remained hidden to Raman and X-ray scattering experiments \cite{Lee2016,Scagliotti1987,Ghosh2021}. CPS resolves both the amplitudes and the phases of the phonons in FePS$_{3}$ by coherently driving them and visualizing their oscillations in real time. Below the N\'eel temperature (T$_\textrm{N}$) of FePS$_{3}$, this elusive spin-lattice interaction manifests itself as a sudden change in the phase and amplitude of the 7.51 THz phonon mode. We found that this interaction is mode-specific, as the 11.45 THz phonon mode does not show any change in its phase or amplitude. In order to pin down the origin of this mode-selective change, we compare and contrast our experimental results with NiPS$_{3}$, which is isostructural to FePS$_{3}$ \cite{Ouvrard1985,Joy1992}. Using crystal field calculations, we find the origin of this effect as a magnetoelastic coupling between $t_{\textrm{2g}}$ orbitals and trigonal distortion, which overlaps strongly with the symmetry of the 7.51 THz A$_\textrm{1g}$ mode. Finally, our analysis identifies the origin of the observed mode-specific spin-phonon coupling in FePS$_{3}$, and hence resolves the dichotomy between low energy and structural collective modes.

In our experimental scheme (Figure 1a), a low intensity and ultrashort pump pulse displacively excites the coherent phonons of FePS$_{3}$ with A$_{1g}$ and E$_{g}$ symmetries without destroying the magnetic order (Methods). After the pump excitation, a lower intensity probe pulse tracks the coherent phonon oscillations as a function of pump-probe delay time $\Delta t$. The center wavelength of the pump is 760 nm (1.63 eV) and the pulse bandwidth is 60 nm. In this energy range, our broadband photoexcitation overlaps with the charge transfer gap \cite{Piacentini1982,Zhang2021}. Figure 1c shows the transient reflectivity trace of FePS$_3$ at room temperature. In addition to the incoherent electronic decay, the signal consists of an oscillatory part composed of two Fourier components (Figure 1c inset), with frequencies of 7.51 THz and 11.45 THz. The lattice distortions corresponding to these phonon modes of A$_{1g}$ symmetry are shown in Figure 1b. The 7.51 THz mode is an out-of-plane breathing mode of sulfur atoms, whereas the 11.45 THz mode involves the in-plane motion of sulfur atoms.

To investigate the effects of spin-lattice coupling on the coherent phonon spectrum, we performed temperature dependent phase-resolved CPS on FePS$_3$ and traced the changes in transient reflectivity as a function of temperature (Figure 2a). Around 90 K, the signal develops a long-lived component, indicating a change in electronic structure due to the magnetic order. Concurrent with this change, we observe a change in coherent phonon oscillations (Figure 2a), which are extracted by subtracting the incoherent background with a single exponential fit. Figure 2b shows the temperature-dependent Fourier transform of the coherent phonon oscillations. Below $\sim$90 K, coherent phonon spectrum develops a new mode at 3.28 THz. This low energy mode has been observed in Raman spectroscopy below T$_\textrm{N}$ and attributed to magnetic zone-folding \cite{Lee2016,Ghosh2021}, heralding the onset of magnetic order. We use this mode as a proxy for magnetic order in our experimental scheme. We attribute the discrepancy between the reported N\'eel temperature (118K) and the observed N\'eel temperature (90 K) to steady-state laser induced average heating in our experiment, due to high repetition rate of our laser. As shown in Figure 2b and 2c, simultaneously with the emergence of the 3.28 THz mode and hence with the onset of the magnetic order, the 7.51 THz mode amplitude shows a clear downturn. Upon cooling, this coherent phonon oscillation vanishes at around 80 K and recovers with further cooling. The 11.45 THz mode, on the other hand, shows negligible change across T$_\textrm{N}$.

We further examine the time-domain evolution of these modes by performing Fourier filtering. The filtered spectral region and the respective time traces are given in Figure 2b. Strikingly, the 7.51 THz mode exhibits a $\pi$ phase shift at low temperatures, which is absent in the 11.45 THz phonon mode, as shown in Figure 2c. Figure 2c also shows the phase-corrected coherent phonon amplitudes, where the $\pi$ phase shift corresponds to negative mode amplitude. The 3.28 THz phonon mode shows an order parameter behaviour, indicating that the pump pulse does not destroy the magnetic order at all temperatures. The 7.51 THz mode amplitude starts to decline at the onset of magnetic order, and its decrease follows the same order parameter behaviour as the 3.28 THz one. This behaviour is strongly suppressed in the 11.45 THz phonon mode, suggesting a smaller magnetoelastic coupling compared to the 7.51 THz mode.

The frequencies of both phonon modes are in agreement with the Raman results. However, in contrast to CPS, the Raman scattering amplitudes for the 7.51 THz and 11.45 THz modes do not show any temperature dependence (SI Figure 2). Despite the sensitivity of both Raman and CPS to phonon modes, both methods excite and detect phonons differently. Raman scattering measures phonon induced changes in polarizability, also known as Raman matrix element ($|\frac{\partial \epsilon}{\partial Q}|^{2}$) by making a transition between the ground state and the single phonon excited state \cite{Loudon1964}. This implies that in FePS$_{3}$, the Raman matrix elements of both A$_{1g}$ phonons remain constant at all temperatures. Contrary to Raman spectroscopy, CPS relies on pump induced coherent phonons generated through displacive or impulsive processes \cite{Zeiger1992}. Thus, the observed mode amplitude in CPS is equal to:

\begin{equation}
    R_{osc} = \sum_{i}\frac{\partial R}{\partial \epsilon}\frac{\partial \epsilon}{\partial Q_{i}}\Delta Q_{i}
\end{equation}

\noindent where $\Delta Q_{i}$ corresponds to the initial displacement of a phonon after photoexcitation, $\epsilon$ is the dielectric constant and $R$ is the reflectivity as a function of dielectric constants. As the Raman matrix elements stay the same at all temperatures, the temperature dependent amplitude and phase changes in CPS can be attributed to the change in $\Delta Q_{i}$.

To gain more insights into the physical origins leading to the mode-selective and magnetic order-dependent phonon phase shift in FePS$_{3}$, we study the free energy landscape before and after the photoexcitation. For ultrafast excitation, the initial phonon displacement $\Delta Q$ equals the difference between the minimal energy lattice positions before and after pump excitation. This mechanism, also known as displacive excitation of coherent phonons, for FePS$_{3}$ is shown in Figure 3, where the upper and lower free energy manifolds represent photoexcited and equilibrium states, respectively. Above T$_\textrm{N}$, the equilibrium and photoexcited free energy landscapes have different minima because of the presence of excited electrons. Below T$_\textrm{N}$, the equilibrium atomic positions shift due to magnetoelastic coupling, and the amount of displacement along specific mode direction depends on the strength of the mode selective spin-phonon coupling. In this case, the pump excitation perturbs the magnetic order and causes the magnetoelastic coupling to relax, leading to an excited free energy manifold similar to the one above T$_\textrm{N}$. If the magnetoelastic coupling is linear in phonon position operators, the magnetic order alters the coherent phonon displacements and phases without changing their frequencies and Raman matrix elements, as shown in Figure 3. The mathematical form of the phenomenological free energy describing this scenario is given in SI. Because the 7.51 THz phonon mode shows a much larger change than the 11.45 THz phonon mode concurrent with the magnetic order, we can explain our experimental findings with the fact that the magnetoelastic coefficient of the 7.51 THz mode is significantly higher than that of the 11.45 THz mode.

Even though our phenomenological model captures our experimental observations with a mode-selective spin-lattice coupling model, it does not provide a microscopic reason for its mode-specificity and the linear coupling between the phonons and magnetic order. To pin down the microscopic origin of these observations, we compare the coherent phonon spectra of FePS$_{3}$ to that of NiPS$_{3}$, which is isostructural to FePS$_{3}$. Both systems develop a zigzag AFM order with similar exchange coupling constants \cite{Lancon2018, Sivadas2015,Kim2021}. In addition to their magnetic structures, the optical spectra reflect similar bandgap for both \cite{Ergecen2022,Kang2020,Piacentini1982}. The significant difference between these two systems are in the electronic configuration of transition metal ions. Ni$^{2+}$ ions (3d$^8$) have two more d-electrons than the Fe$^{2+}$ ions (3d$^6$). Although the CPS spectra of NiPS$_{3}$ (see Figure S3) exhibit both 7.51 and 11.45 THz phonon modes with the same symmetry as FePS$_{3}$, none of these modes show any temperature-dependent phase or amplitude change below T$_\textrm{N}$. Therefore, we can ascribe the mode-selective spin-lattice coupling in FePS$_{3}$ to its localized d-orbital electron configuration.

The d-orbital electron configurations of Ni$^{2+}$  and Fe$^{2+}$ ions are given Figure 4a. In both compounds, transition metal ions are surrounded by ligands with octahedral arrangements, together with trigonal distortions \cite{Joy1992}. Both Fe$^{2+}$ and Ni$^{2+}$ ions have two unpaired spins in $e_{\textrm{g}}$ orbitals. However, $t_{\textrm{2g}}$ levels in Fe$^{2+}$ are partially-filled, and in Ni$^{2+}$, these levels are fully-filled. Therefore we can narrow down the microscopic origin of the mode-selective magnetoelasticity in FePS$_{3}$ to $t_{\textrm{2g}}$ orbitals.

To examine the magnetoelastic effects in FePS$_{3}$, we theoretically analyze the system on a single octahedra level and focus on the low energy $t_{\textrm{2g}}$ manifold. In the presence of trigonal distortions, the Hamiltonian for the low energy $t_{\textrm{2g}}$ manifold can be written as \cite{Joy1992}:

\begin{equation}
    H = (\Delta_{\textrm{trig.}} + \alpha u) L^2_{z}+ \lambda L S + \frac{1}{2}Bu^2
\end{equation}

\noindent where $\Delta_{\textrm{trig.}}$ is the existing trigonal splitting, $\lambda$ is the spin-orbit coupling constant, $B$ is the elastic constant associated with trigonal distortions, and $\alpha$ quantifies the change in energy following a change in trigonal distortions. The quantization axis is along the (111) direction of the octahedra and the c-axis of the crystal. Despite the absence of a direct coupling between the structural distortion $u$ and the spin operator $S$, spin-orbit coupling gives rise to an effective Hamiltonian, that is proportional to $H_{\textrm{eff, magnetoelastic}} = - \frac{\alpha \lambda^2}{\Delta^2} u S^2_{z}$. This term implies that the z-component of spins selectively displaces the octahedra along the trigonal distortion. This does not cause any change in its elastic properties, which would appear as a coupling quadratic term in $u$ and would alter the normal mode frequencies. Furthermore, this microscopic treatment yields the value of trigonal distortion as $u_0 = \frac{\alpha\lambda^2}{B\Delta^2}S_z^2$. Following the pump excitation, this value will get altered because of the pump induced perturbation of the magnetic system, which is equal to $\delta u_0 = \frac{2\alpha\lambda^2}{B\Delta^2}S_{z} \delta S_{z}$. This expression is identical to the magnetic order-dependent phonon displacement ($\Delta Q$) expression and explains the spin-dependent coherent phonon oscillations microscopically on a single ion level. 

Our microscopic analysis of mode-selective magnetoelasticity that relies on single ion anisotropy is valid for FePS$_{3}$ on the bulk level. First, below the magnetic transition, the expectation value of $S^2_z$ is constant for different octahedra composing the material. Therefore, each individual layer of FePS$_{3}$ undergoes a uniform trigonal distortion, and $S_z$ operator can be replaced with the antiferromagnetic order parameter $N$. Furthermore, the spin-orbit coupling in this material emerges from on-site $e^{\pi}_g$ d-orbitals. Unlike other vdW magnets such as CrI$_{3}$ \cite{Kim2019, Lado2017}, the spin-orbit coupling does not emerge from ligands, rendering our single-ion treatment valid for FePS$_{3}$ on the bulk level. 

The trigonal distortions induced by the mode-selective magnetoelasticity on the single octahedra level correspond to out-of-plane motion of sulfur atoms on the bulk level. Therefore, the mode-selective magnetoelasticity will couple to phonon modes differently, depending on their spatial projection on the trigonal distortion or out-of-plane motion of sulfur atoms. As shown in Figure 1, 7.51 THz A$_{1g}$ phonon mode modulates the out of plane distance of sulfur atoms and directly corresponds to the trigonal distortion of octahedra. On the other hand, as the 11.45 THz phonon mode involves in-plane motion of sulfur atoms, the coupling between this phonon mode and the trigonal distortions is negligible. Furthermore, as shown in Figure 4b, this magnetoelastic coupling and phonon phase shift are absent in NiPS$_{3}$ because of the fully filled t$_{2g}$ orbitals.

In summary, using coherent phonon spectroscopy, we reveal mode selective magnetoelasticity in FePS$_{3}$, which previously eluded phase insensitive measurements. This effect changes the coherent phonon amplitude and the phase of the 7.51 THz A$_{1g}$ phonon mode without frequency renormalization below the magnetic transition temperature. By performing a comparative study between FePS$_{3}$ and NiPS$_{3}$, we pinpoint the mode selective origin of spin-lattice coupling and show the pivotal role of trigonal distortions in these compounds. Our results not only reveal a mode selective magnetoelasticity in FePS$_{3}$ but resolve the dichotomy between magnetically enabled phonon modes and already existing phonon modes. Our results suggest that perturbations that directly couple to the trigonal distortions in FePS$_{3}$, such as pressure \cite{Coak2021} and nonlinear phononics \cite{Disa2020, Fiete2021}, can be used to manipulate the magnetic order or to enter nonequilibrium magnetic phases which cannot be accessed in equilibrium. Furthermore, we envision that the coherent phonon spectroscopy technique can be utilized as a highly sensitive probe of hidden spin-lattice coupling in vdW magnet monolayers and other systems with strong spin-lattice coupling.

\begin{figure*}
\centering
\includegraphics[scale=0.8]{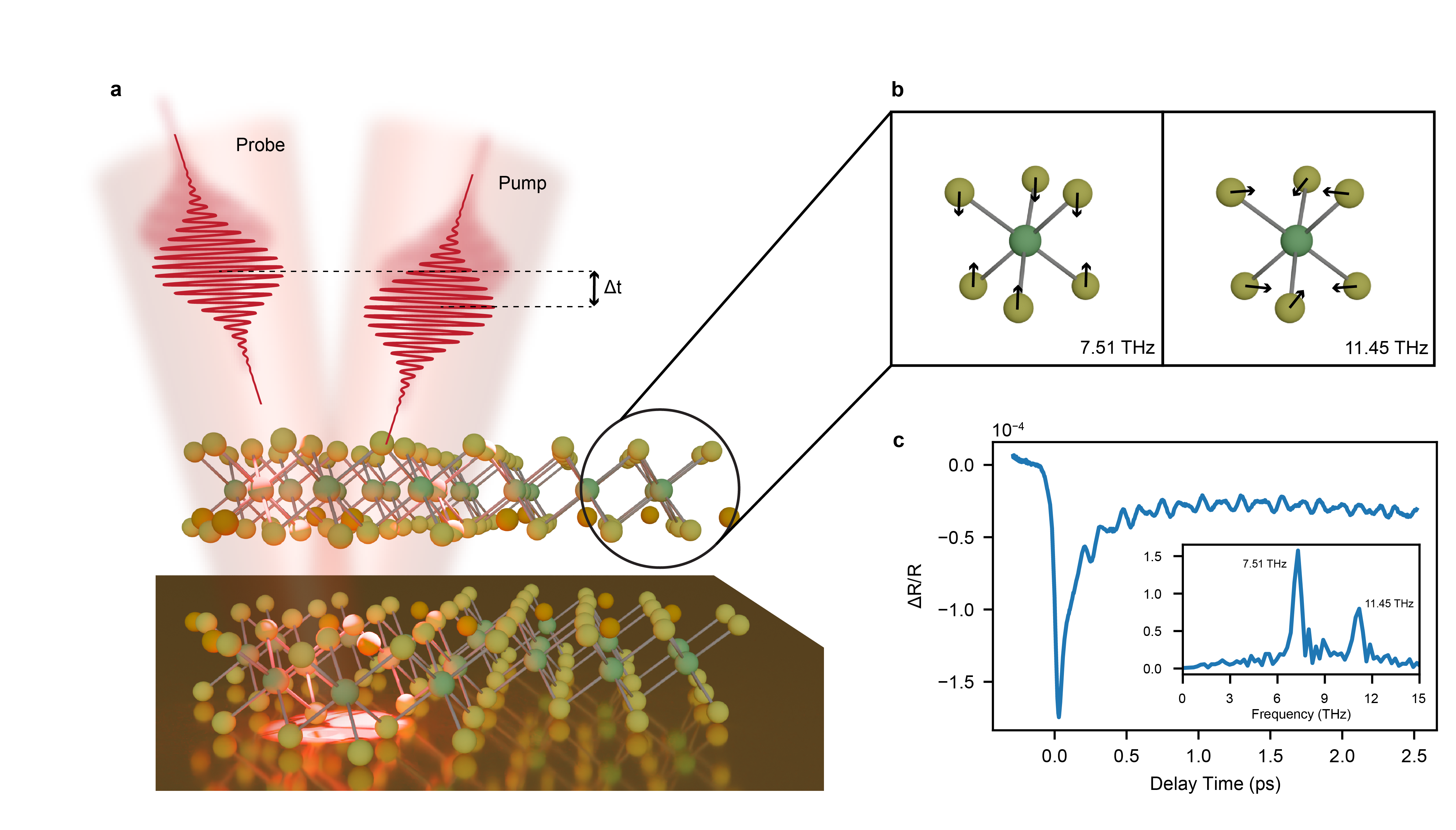}
\caption{\textbf{Coherent phonon spectroscopy on FePS$_3$.} \textbf{a.} An ultrashort pump pulse of 1.63 eV photon energy excites the crystal, and a subsequent probe pulse measures the transient differential reflectivity after a delay time of $\Delta t$. \textbf{b.} In FePS$_3$, six sulfur (yellow) atoms in an octahedral configuration surround iron (green) atoms. Arrows depict the octahedral distortions that correspond to two $A_\textrm{{1g}}$ phonons with frequencies 7.51 THz and 11.45 THz. \textbf{c} The transient reflectivity trace of FePS$_{3}$ obtained at room temperature (293 K) consists of an incoherent electronic decay signal and oscillatory features originating from coherent phonons. The Fourier spectrum (inset) of these oscillatory features reveals two phonon modes with frequencies matching the $A_\textrm{{1g}}$ phonons shown in \textbf{b}.}
\end{figure*}

\begin{figure*}
\centering
\includegraphics[scale=0.95]{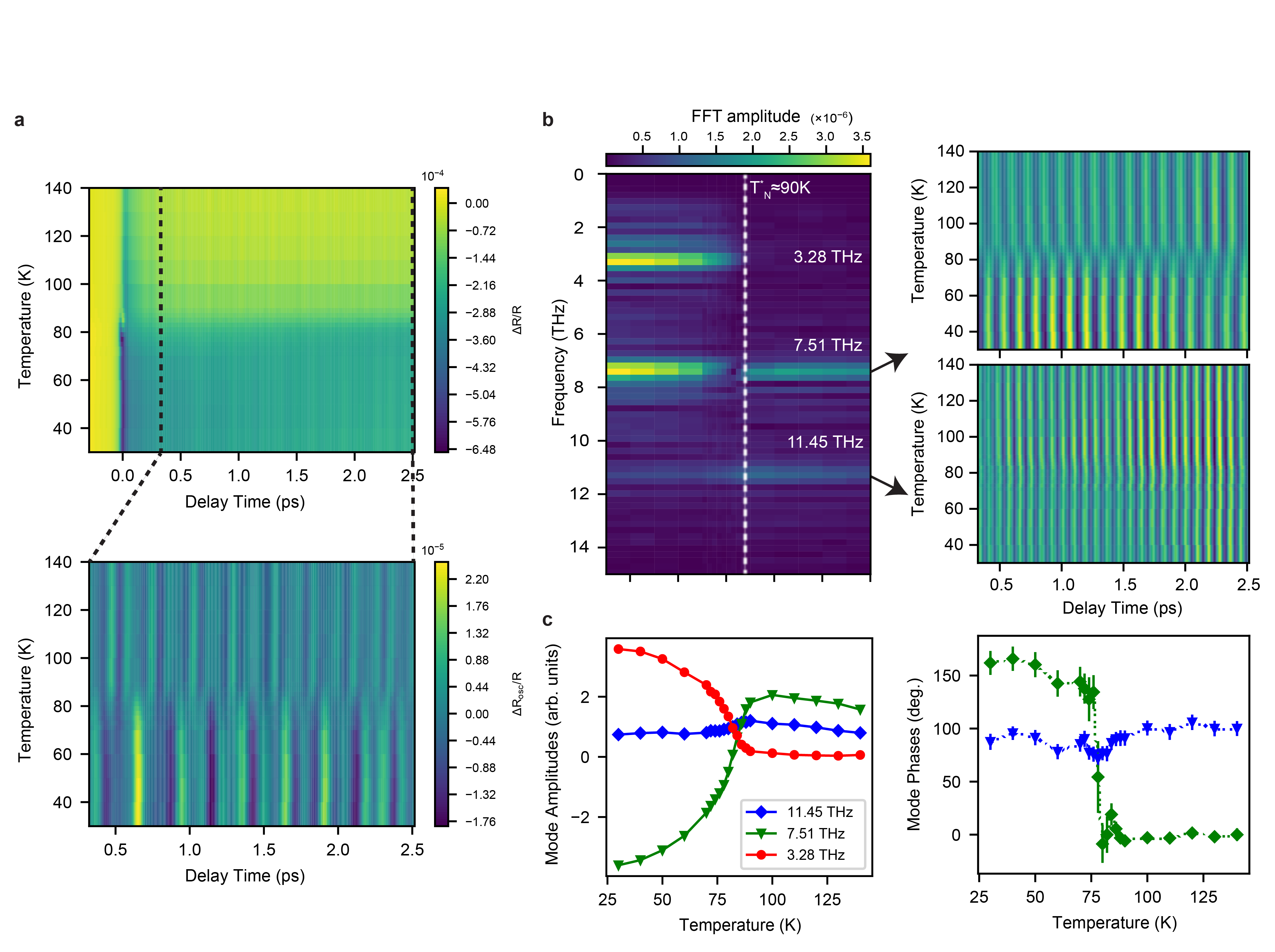}
\caption{\textbf{The N\'eel order selectively couples to the 7.51 THz $A_\textrm{{1g}}$ phonon mode.} \textbf{a.} The temperature dependent coherent phonon spectroscopy on FePS$_3$ exhibits a pronounced change near $\sim$90 K (upper panel). Lower panel shows the oscillatory part of the traces, extracted by fitting the incoherent background with a single exponential. \textbf{b.} (Left panel) Below $\sim$90 K, in temperature dependent Fourier spectra of oscillations, a new phonon emerges mode at 3.28 THz. Fourier filtered time traces of the 7.51 THz and 11.45 THz modes are shown on the right. \textbf{c.} Temperature dependent phase-corrected amplitudes and phases of phonons. The 3.28 THz mode displays an order parameter-like behavior, and its onset is concomitant with the magnetic order. Following the emergence of this mode, the 7.51 THz mode undergoes a $\pi$-phase shift and hence its phase-corrected amplitude changes sign. The 11.45 THz mode amplitude shows no discernable change.}
\end{figure*}  

\begin{figure}
\centering
\includegraphics[scale=0.9]{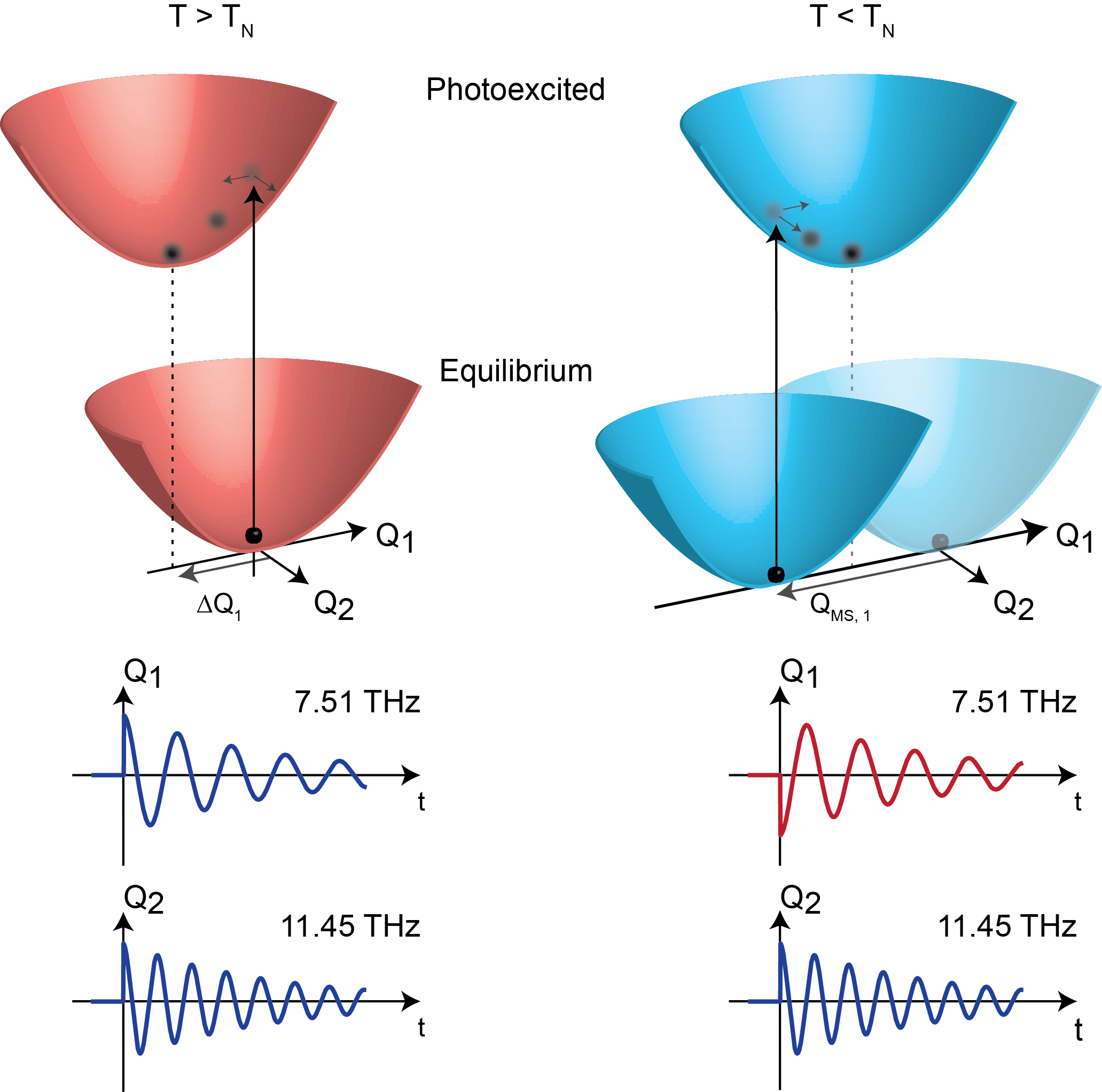}
\caption{\textbf{Magnetoelasticity displaces the equilibrium energy landscape anisotropically and induces a mode-selective $\pi$-phase shift in coherent phonon oscillations}. Upper panels show the equilibrium and photoexcited free energy landscapes. Q$_1$ and Q$_2$ are generalized atomic coordinates along the motion axis of 7.51 THz and 11.45 THz phonons, respectively. Above T$_\textrm{N}$, photoexcitation launches coherent phonons displacively due to a relative shift between photoexcited and equilibrium free energy landscape minima (e.g $\Delta$Q$_1$ along the Q$_1$ axis). Below T$_\textrm{N}$, magnetoelasticity induces an anisotropic shift of the equilibrium landscape, which is much stronger along the Q$_1$ axis (Q$_{\textrm{MS, 1}}$) than Q$_2$ axis. This results in $\pi$-phase shift of the 7.51 THz mode, while 11.45 THz mode remains unaffected. This scenario is depicted in the lower panels which show the coherent oscillations of each mode in time domain below and above T$_\textrm{N}$.}
\end{figure}

\begin{figure}
\centering
\includegraphics[scale=0.9]{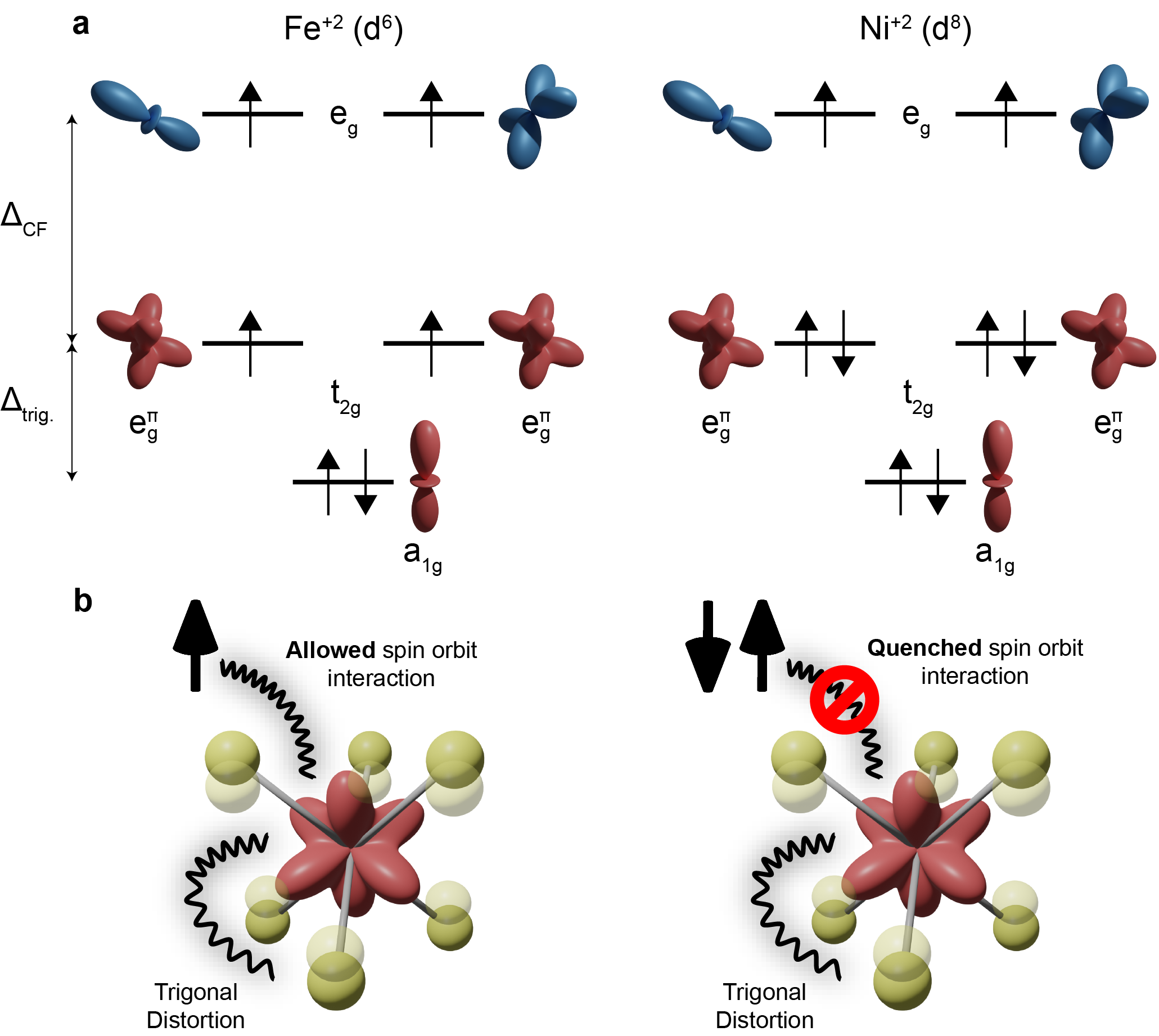}
\caption{\textbf{Spins in $e^\pi_\textrm{{g}}$ levels selectively couple to trigonal distortions and give rise to mode-selective magnetoelasticity in FePS$_3$}. \textbf{a.} Electronic configurations of Fe$^{2+}$ and Ni$^{2+}$  ions in a trigonally contracted octahedral environment of sulfur atoms. In FePS$_3$, transition metal ions host two unpaired spins in $e^\pi_\textrm{{g}}$ orbitals and two unpaired spins in $e_\textrm{{g}}$ orbitals with quenched orbital angular momenta. On the other hand, in NiPS$_3$, transition metal ions host two additional electrons in $e^\pi_\textrm{{g}}$ orbitals compared to FePS$_3$. \textbf{b.} Two unpaired spins in $e^\pi_\textrm{{g}}$ orbitals of FePS$_3$ allow efficient coupling between the magnetic order and trigonal distortions. This magnetoelastic effect is due to spin-orbit coupling mediated virtual transitions between $e^\pi_\textrm{{g}}$ and $a_\textrm{{1g}}$ orbitals. This effect is absent in NiPS$_3$ (shown on the right), since the $t_\textrm{{2g}}$ levels are fully filled.}
\end{figure}  

\matmethods{\subsection*{Sample preparation}
We synthesized our FePS$_3$ crystals using a chemical vapor transport method (for details see Ref. \cite{Kuo2016}). All the powdered elements (purchased from Sigma-Aldrich): iron (99.99\% purity), phosphorus (99.99\%) and sulfur (99.998\%), were prepared inside an argon-filled glove box. After weighing the starting materials in the correct stoichiometric ratio, we added an additional 5 wt of sulfur to compensate for its high vapor pressure. After the synthesis, we carried out the chemical analysis of the single-crystal samples using a COXI EM-30 scanning electron microscope equipped with a Bruker QUANTAX 70 energy dispersive X-ray system to confirm the correct stoichiometry. We also checked the XRD using a commercial diffractometer (Rigaku Miniflex II). Prior to optical measurements, we determined the crystal axes of the samples using an x-ray diffractometer. We cleaved samples before placing them into high vacuum ($\sim10^{-7}$ torr) to expose a fresh surface without contamination and oxidation.

\subsection*{Phase-resolved coherent phonon spectroscopy}
A Ti:sapphire oscillator (Cascade-5, KMLabs), centered at 760 nm (1.63 eV) and with pulse duration of $\sim$25 fs was used in our experiments. The repetition rate of the laser was set to 80 MHz. Before splitting the output into pump and probe branches, we compensated for group velocity dispersion (GVD) using a pair of chirp mirrors and N-BK7 wedges to maintain the pulse duration at the sample position. The pump and probe pulses were characterized separately at the sample position, using frequency resolved optical gating technique. The pulse duration was $\sim$25 fs. To increase the signal to noise ratio of our setup, we modulate the pump intensity at 100 kHz. The probe signal from the photodiode is sent to a lock-in amplifier (Stanford Research SR830) locked to the chopping frequency (100 kHz). For faster data acquisition and averaging, the pump-probe delay is rapidly scanned at a rate of 5 Hz with an oscillating mirror (APE ScanDelay USB). The diameters of pump and probe beam spots were 90 $\mu$m, measured by a knife edge method. The pump and probe beams are cross polarized. During all of our measurements, the pump fluence was set to 10 $\mu J / cm^2$. The detailed schematics of the setup is given in SI (Figure S1).

\subsection*{Data, Materials, and Software Availability}
All study data are included in the article and/or supporting information.
}

\showmatmethods{} 

\acknow{We thank Riccardo Comin for fruitful discussions. We acknowledge support from the US Department of Energy, BES DMSE (data taking and analysis) and Gordon and Betty Moore Foundation's EPiQS Initiative grant GBMF9459 (instrumentation and manuscript writing). Work at the Center for Quantum Materials was supported by the Leading Researcher Program of the National Research Foundation of Korea (Grant No. 2020R1A3B2079375). The research of S.O. was supported by
the U.S. Department of Energy, Office of Science, Basic Energy Sciences, Materials Sciences and Engineering Division.}

\showacknow{} 

\end{document}